\documentclass[twocolumn]{aastex631}

\usepackage{ifthen}
\usepackage{afterpage}

\newcommand{\comment}[1]{}

\usepackage{amsmath}
\usepackage{natbib}
\usepackage{appendix}
\usepackage{gensymb}
\usepackage{wasysym}
\usepackage{bm}

\providecommand{\adsurl}[1]{\href{#1}{ADS}}

\newcommand{\revision}[1]{#1}
\DeclareSymbolFont{UPM}{U}{eur}{m}{n}
\DeclareMathSymbol{\umu}{0}{UPM}{"16}
\let\oldumu=\umu
\renewcommand\umu{\ifmmode\oldumu\else$\oldumu$\fi}
\newcommand\micro{\umu}

\newcommand\microns{\micro m}

\comment{\hfill\textbf{DRAFT of {\today} \now}.}

\shorttitle{Eclipse Mapping Null Space}
\shortauthors{Challener \& Rauscher}

\begin{document}

\title{The Eclipse Mapping Null Space: Comparing Theoretical Predictions with Observed Maps}
\author[0000-0002-8211-6538]{Ryan C. Challener}
\affiliation{Department of Astronomy, University of Michigan, 1085
  S. University Ave., Ann Arbor, MI 48109, USA}

\author[0000-0003-3963-9672]{Emily Rauscher}
\affiliation{Department of Astronomy, University of Michigan, 1085
  S. University Ave., Ann Arbor, MI 48109, USA}


\begin{abstract}

High-precision exoplanet eclipse light curves, like those possible with JWST, enable flux and temperature mapping of exoplanet atmospheres.
These eclipse maps will have unprecedented precision, providing an opportunity to constrain current theoretical predictions of exoplanet atmospheres. 
However, eclipse mapping has unavoidable mathematical limitations because many map patterns are unobservable.
This ``null space'' has implications for making comparisons between predictions from general circulation models (GCMs) and the observed planet maps, and, thus, affects our understanding of the physical processes driving the observed maps.
We describe the eclipse-mapping null space and show how GCM forward models can be transformed to their observable modes for more appropriate comparison with retrieved eclipse maps, demonstrated with applications to synthetic data of an ultra-hot Jupiter and a cloudy warm Jupiter under \revision{JWST-best-case-} and extreme-precision observing scenarios.
We show that the effects of the null space can be mitigated and manipulated through observational design, and JWST exposure times are short enough to not increase the size of the null space.
Furthermore, we show the mathematical connection between the null space and the ``eigenmapping'' method, demonstrating how eigenmaps can be used to understand the null space in a model-independent way.
We leverage this connection to \revision{incorporate null-space uncertainties in retrieved maps, which increases the uncertainties to now encompass the ground truth for synthetic data.}
The comparisons between observed maps and forward models that are enabled by this work, and the improved eclipse-mapping uncertainties, will be critical to our interpretation of multidimensional aspects of exoplanets in the JWST era.


\end{abstract}


\section{INTRODUCTION}
\label{sec:introduction}

Exoplanet eclipses enable mapping of the planet's surface or photosphere.
As the planet passes behind its host star, the brightness of the system changes based on the flux emitted by the regions of the planet being covered or uncovered by the stellar disk \citep{WilliamsEtal2006apjEclipseMapping, RauscherEtal2007apjEclipseMapping}.
Thus, the light curve of the planet-star system can be inverted to retrieve a brightness map of the planet, and, in turn, a map of the temperatures of the planet. 
This has been done successfully for HD 189733 b by stacking several Spitzer Space Telescope 8 \microns\ eclipse observations \citep{DeWitEtal2012aaHD189Map, MajeauEtal2012apjlHD189Map, RauscherEtal2018ajMap, ChallenerRauscher2022ajThERESA} and was recently done with a JWST NIRISS/SOSS eclipse of WASP-18b \citep{CoulombeEtal2023arxivWASP18b}.
Now that JWST is operational, many more hot and ultra-hot Jupiters are mappable.

Hot Jupiters are a class of planet where General Circulation Models (GCMs) predict large temperature gradients between the permanent day- and night-sides \citep[e.g.,][]{ShowmanGuillot2002aapGCM, ShowmanEtal2009apjHD189HD209GCM, RauscherMenou2012apjGCM, HengEtal2011mnrasDoubleGrayGCM, Dobbs-DixonAgol2013mnrasHD189GCM, MayneEtal2014aapGCM, ChoEtal2015mnrasGCM}, although their detailed three-dimensional structures will depend on the atmospheric wind patterns and how they may be shaped by additional processes such as chemistry, clouds or hazes, magnetic effects, and hydrogen dissociation \citep[e.g.,][]{RogersShowman2014apjlHD209MHD, HellingEtal2016mnrasHD209HD189clouds, KatariaEtal2016apjCircChemSurvey, LeeEtal2016aapHD189CloudsI, PartmentierEtal2016apjCloudTransitions, DrummondEtal2018apjHD1893D, TanKomacek2019apjUHJcirculation, RomanEtal2021apj3DClouds, SteinrueckEtal2021mnras3DHazeHD189, BeltzEtal2022ajActiveMagDrag3D}.
However, observational constraints on exoplanet atmosphere have been limited to hemispherically-averaged properties due to data quality limitations \citep[e.g.,][]{StevensonEtal2017ajWASP43bPhaseCurve, ArcangeliEtal2019aapW18PhaseCurve, MayEtal2021ajWASP76bPhaseCurves, MayEtal2022ajSpitzerPhaseCurves}.

JWST provides the first opportunity to understand exoplanet atmospheres on a three-dimensional level. 
Eclipse maps will provide the first empirical challenges for multidimensional aspects of GCMs.
For example, the first JWST eclipse map is of an ultra-hot Jupiter; from the sharp drop in brightness near the terminator, GCMs indicate that a strong (possibly magnetic) drag mechanism must be at work, preventing winds from more efficiently advecting hot gas away from the dayside \citep{CoulombeEtal2023arxivWASP18b}.
Furthermore, GCMs will be used for physical interpretations of eclipse map features like shifted hotspot locations and temperature gradients. 
Thus, comparisons between observed planet maps and theoretical predictions will be critical to our understanding of exoplanet atmospheres through the JWST era and beyond.
Understanding the nuance of these comparisons is crucial.

Not all brightness patterns on a planet are mappable \citep{CowanEtal2013mnrasMapping}. 
This is intuitive for rotational light curves (planetary phase curves); we can only measure hemispherically-integrated planet flux as the object rotates, so many patterns have no observable signal \citep{LugerEtal2021ajNullSpace}.
For example, a planet with no variation besides latitudinal bands would appear indistinguishable from a planet with a uniform brightness temperature.
These immeasurable structures, or the ``null space'', persist even in the limit of infinite signal-to-noise (S/N).
While the null space is much smaller if the planet is eclipsed by its host star, it is not empty.

In this paper, we discuss the ramifications of the null space on eclipse mapping and how to best compare GCMs against observed planet maps. 
In Section \ref{sec:null} we give an overview of the null space, in Section \ref{sec:eigen} we show the relationship between the null space and eclipse-mapping models, in Section \ref{sec:mit} we discuss strategies to minimize null-space effects, in Section \ref{sec:gcm} we describe how to transform 3D model predictions to their observable forms, in Section \ref{sec:application} we demonstrate with applications to synthetic observations, in Section \ref{sec:unc} we introduce a method to incorporate the null-space uncertainty into retrieved maps, and in Section \ref{sec:conclusions} we lay out our conclusions.

\section{THE NULL SPACE}
\label{sec:null}

We refer the reader to \cite{LugerEtal2021ajNullSpace} (hereafter L2021), which lays out a thorough description of the null space.
Here, we give a brief overview.

The spherical harmonic functions $Y^l_m(\theta, \phi)$ form an orthogonal basis on a sphere and, when combined with amplitudes $\bm{y} = \{y^0_0, y^1_{-1}, y^1_0, ...\}$, can represent any flux map given harmonics of sufficient degree.
For a maximum spherical harmonic degree $l_{\rm max}$, the flux map is 

\begin{equation}
\label{eqn:map}
    Z(\theta, \phi) = \sum_{l=0}^{l_{\rm max}}\sum_{m=-l}^{l} y_m^l Y_m^l(\theta, \phi),
\end{equation}

\noindent
where $\theta$ and $\phi$ are the latitude and longitude on the sphere.

We note that Equation \ref{eqn:map} can be rewritten as

\begin{equation}
    Z(\theta, \phi) = y^0_0 Y^0_0 + \sum_{l=1}^{l_{\rm max}}\sum_{m=-l}^{l} y_m^l Y_m^l(\theta, \phi),
\end{equation}

\noindent
where the first term is a uniform map with weight $y_0^0$ and the second term is a set of variations on that uniform map.
There are $N = (l_{\rm max} + 1)^2 - 1$ spherical harmonic amplitudes in the second term (the number of spherical harmonics with $l \leq l_{\rm max}$ minus $y^0_0$).
We write the equation in this way to make comparison between the spherical harmonic basis described here and the ``eigenmap'' basis, which we discuss further in Section \ref{sec:eigen}.

From the flux map $Z$, we obtain the planet's observed light curve (with $K$ measurements in time) from

\begin{equation}
\label{eqn:designmatrix}
    \bm{f} = y^0_0 F^0_0(t) + \bm{\mathcal{A}\ y}
\end{equation}

\noindent
where $\bm{\mathcal{A}}$ is the design matrix\footnote{The design matrix is a mathematical representation of the integration of each spherical harmonic map component weighted by the visibility function at each point in time \citep{CowanEtal2013mnrasMapping}.}, a $K \times (N - 1)$ matrix that describes how each spherical harmonic mode contributes to each flux measurement in the light curve, $F^0_0$ is the light curve of the uniform map $Y^0_0$, and $\bm{y}$ is the vector of $y^l_m$ weights with $l > 0$.
We note that the design matrix used in L2021 and our $\bm{\mathcal{A}}$ are slightly different, in that our design matrix does not include $F^0_0$.

\begin{figure*}
    \centering
    \includegraphics[width=7in]{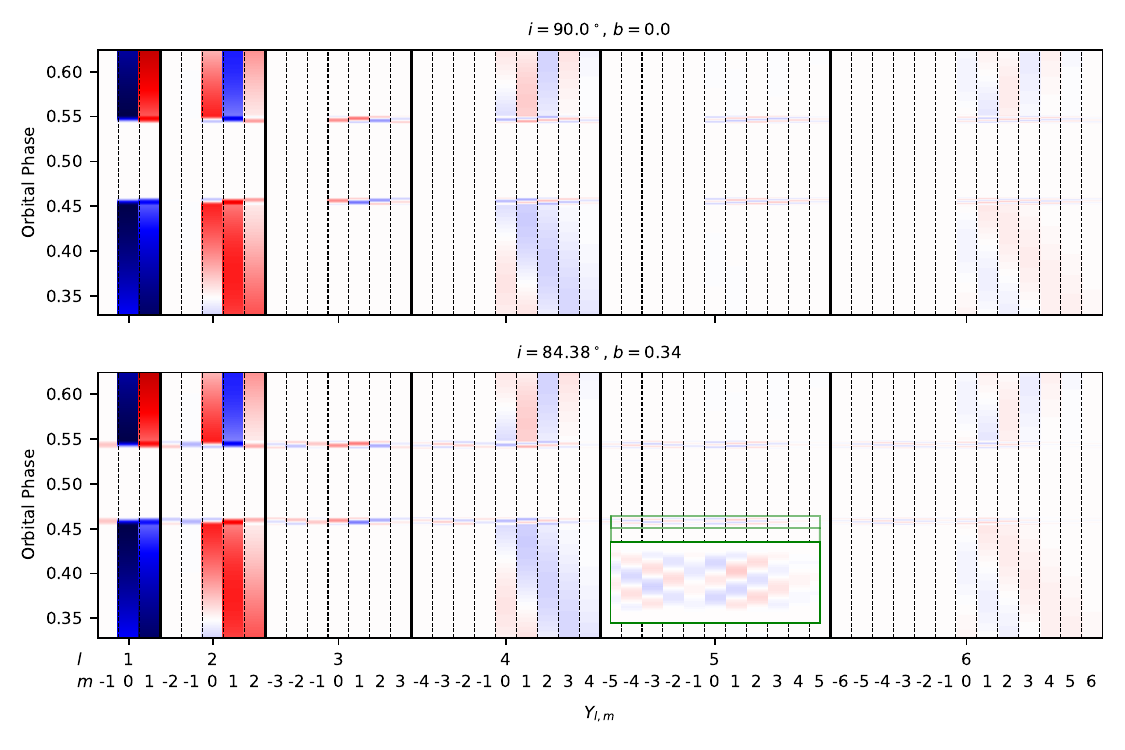}
    \caption{Design matrices for the eclipse observation of WASP-18b presented in \cite{CoulombeEtal2023arxivWASP18b}, shown out to sixth degree harmonics, for a perfectly edge-on (top) and inclined (bottom) orbit. Each column shows the contribution from a given spherical harmonic mode at each time (or orbital phase) in the observation. Multiplying a row by harmonic weights $\bm{y}$, summing, and adding the uniform component gives the planetary light curve at that time (Equation \ref{eqn:designmatrix}). The vertical black lines separate the harmonic degrees. Red and blue indicate positive and negative contributions to the light curve, respectively. The white regions show times where a harmonic contributes zero flux (i.e., that harmonic is unobservable at that time). The green inset zooms in on examples of the spherical harmonic signals generated in the eclipse ingress.}
    \label{fig:designmatrix}
\end{figure*}

Figure \ref{fig:designmatrix} shows the design matrix out to sixth-degree harmonics for an observation matching that in \cite{CoulombeEtal2023arxivWASP18b}.
Outside of eclipse, ingress, and egress, the contribution to the light curve is zero for many harmonics, because many modes are not measureable with rotational light curves (L2021, \citealp{CowanEtal2013mnrasMapping}).
If the object is eclipsed, harmonic modes contribute information to the design matrix during eclipse ingress and egress, so more spatial information becomes available, as the eclipse breaks many latitudinal degeneracies. 
Furthermore, if the planet's orbit has an inclination $i < 90^\circ$ (impact parameter $b > 0$), latitudinally-symmetric modes become observable\footnote{Generally there is ambiguity about the orientation of the inclination. In this work, we assume the planets rotate and revolve about the same axis, and that the planets eclipse north of the stellar equator.}.

\begin{figure*}[t]
    \centering
    \includegraphics[width=7in]{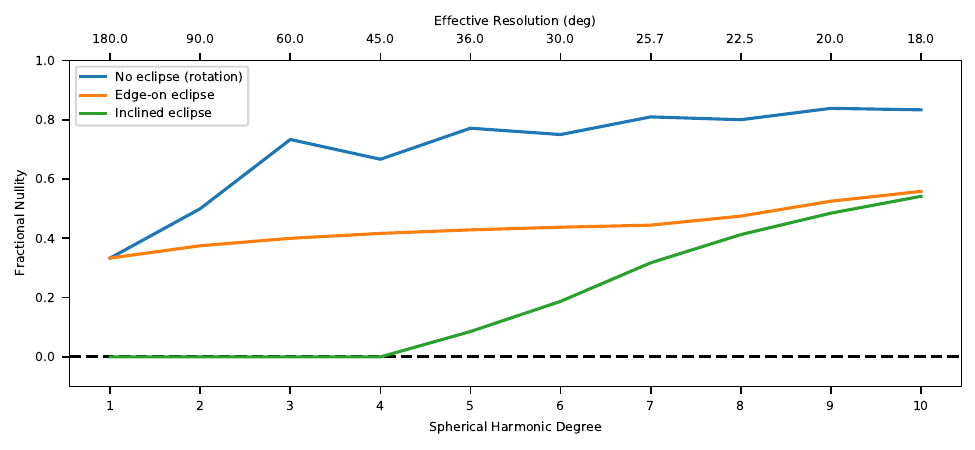}
    \caption{The nullity (how much spatial information is unmappable) for a rotational light curve vs.\ an eclipsing light curve. We assume the inclined eclipsed object has an orbital inclination of $89^\circ$ to break degeneracies of latitudinally-symmetric spherical harmonic modes. The eclipse gives an observation access to significantly more brightness patterns on the planet than for a non-eclipse planet, particularly if the orbit is inclined. However, many patterns remain unobservable, especially at high spatial resolution (high spherical harmonic degree). \revision{In all cases the planet-to-star radius ratio is 0.1, mass ratio is 0.001, the orbital period is 1 day, and the full planet orbit is sampled 100,000 times such that sampling rate does not impact the nullity at these spatial resolutions. See later sections for further discussion of the effects of orbital parameters and sampling rate. We make no assumption about the signal-to-noise ratio of the observation as it has no effect on the nullity.}}
    \label{fig:nullity}
\end{figure*}

Although Figure \ref{fig:designmatrix} shows that all spherical harmonics have an eclipse signal for a planet on an inclined orbit (i.e., no column is all zeros), many of these signals are degenerate.
The size of the null space (the ``nullity") is $N - R$, where $R$ is the rank of the design matrix $\bm{\mathcal{A}}$.
$R$ is the number of non-degenerate signals in the design matrix and so the nullity is the number of degenerate signals lost to the null space.
Therefore, $(N-R)/N$ is the fraction of information available at a given harmonic degree that is inaccessible to an observation.
Figure \ref{fig:nullity} shows this fractional nullity for a non-eclipsing object, an eclipsing object on an edge-on orbit, and an eclipsing object with an inclined orbit, for spherical harmonic degree $\le10$ as a fraction of the total number of modes.
An inclined eclipse breaks most of the degeneracies for spherical harmonic degree $\le4$, but many of the higher-order harmonics with finer spatial structure remain degenerate and, thus, are lost to the null space.

L2021 showed that $\bm{\mathcal{A}}$ can be decomposed into two operators, $\bm{P}$ and $\bm{N}$, which separate out the components of $\bm{y}$ that affect the light curve and those which do not, using singular value decomposition (SVD). 
Briefly, the design matrix can be written as 

\begin{equation}
\label{eqn:svd}
    \bm{\mathcal{A}} = \bm{U\ S\ V^T}
\end{equation}

\noindent
where $\bm{U}$ is a $K \times K$ orthogonal matrix, $\bm{S}$ is a $K \times N$ diagonal matrix, and $\bm{V}$ is an $N \times N$ orthogonal matrix.
These matrices can be split into observable ($\bm{\bullet}$) and null ($\bm{\circ}$) components based on the rank $R$ of $\bm{\mathcal{A}}$ such that

\begin{equation}
    \bm{U} \equiv (\bm{U_\bullet} | \bm{U_\circ})
\end{equation}

\begin{equation}
    \bm{S} \equiv \left(\frac{\bm{S_\bullet}}{\bm{0}} | \frac{\bm{0}}{\bm{S_\circ}}\right)
\end{equation}

\begin{equation}
    \bm{V^T} \equiv \left(\frac{\bm{V^T_\bullet}}{\bm{V^T_\circ}}\right)
\end{equation}

\noindent
Combining Equations \ref{eqn:designmatrix} and \ref{eqn:svd}, it can be shown that we can define

\begin{equation}
    \bm{\mathcal{P}} \equiv \bm{V_\bullet\ V^T_\bullet}
\end{equation}

\begin{equation}
    \bm{\mathcal{N}} \equiv \bm{V_\circ\ V^T_\circ}
\end{equation}

\noindent
such that

\begin{equation}
    \bm{y_\bullet} = \bm{\mathcal{P}\ y}
\end{equation}

\begin{equation}
    \bm{y_\circ} = \bm{\mathcal{N}\ y},
\end{equation}

\noindent
where $\bm{y_\bullet}$ is the linear combination of spherical harmonic components that contribute to the light curve and $\bm{y_\circ}$ is the linear combination of components which is in the null space. 
We refer the reader to the appendix of L2021 and the associated supplemental materials\footnote{github.com/rodluger/mapping\_stellar\_surfaces} for a thorough derivation and practical examples.

Given a map that is constructed from spherical harmonics, these sets of spherical harmonic weights can be used to construct new maps following Equation \ref{eqn:map}. 
Maps constructed from $\bm{y_\bullet}$ and $\bm{y_\circ}$ are physical representations of the observable modes $Z_{\bullet}$ (referred to as the ``preimage'' in L2021) and the null modes $Z_\circ$.
Note that $Z = Z_{\bullet} + Z_\circ$ and $\bm{f} = \bm{f_\bullet} + \bm{f_\circ}$.
In fact, $\bm{f_\circ} = \bm{0}$, which means that $Z_{\circ}$ can be multiplied by any coefficient before calculating $Z$ without affecting the observed light curve. 
There is no unique solution when inverting eclipse light curves into planet maps.

\section{EIGENMAPPING AND THE NULL SPACE}
\label{sec:eigen}

``Eigenmapping'' is currently the state-of-the-art in eclipse mapping methods \citep{RauscherEtal2018ajMap, MansfieldEtal2020mnrasEigenspectraMapping, ChallenerRauscher2022ajThERESA}.
In brief, this method computes an initial basis set of spherical-harmonic light curves (a design matrix $\bm{\mathcal{A}}$), transforms them to an orthogonal basis of ``eigencurves'' using SVD, and fits the data as a weighted sum of multiple eigencurves, a uniform-map light curve, and a normalization correction term \citep{RauscherEtal2018ajMap}. 
The resulting brightness map of the planet is constructed the same way, as a weighted sum of the ``eigenmaps'' which correspond to each eigencurve and the uniform component.

The eigencurves are constructed through a similar process to the construction of the observable $\bm{P}$ and null $\bm{N}$ operators and are thus related.
In fact, the set of eigencurves, which constitutes a new design matrix $\bm{\mathcal{A}}_{\rm E}$, is given by

\begin{equation}
\label{eqn:ae}
    \bm{\mathcal{A}}_{\rm E} = \bm{U}\ \bm{S} = \bm{\mathcal{A}}\ \bm{V}.\
\end{equation}

\noindent
In other words, the columns of $\bm{V}$ are the eigenvectors of $\bm{\mathcal{A}}$ that project the spherical harmonic light curves into new axes, generating the orthogonal eigencurves.

\begin{figure*}
    \includegraphics[width=\textwidth]{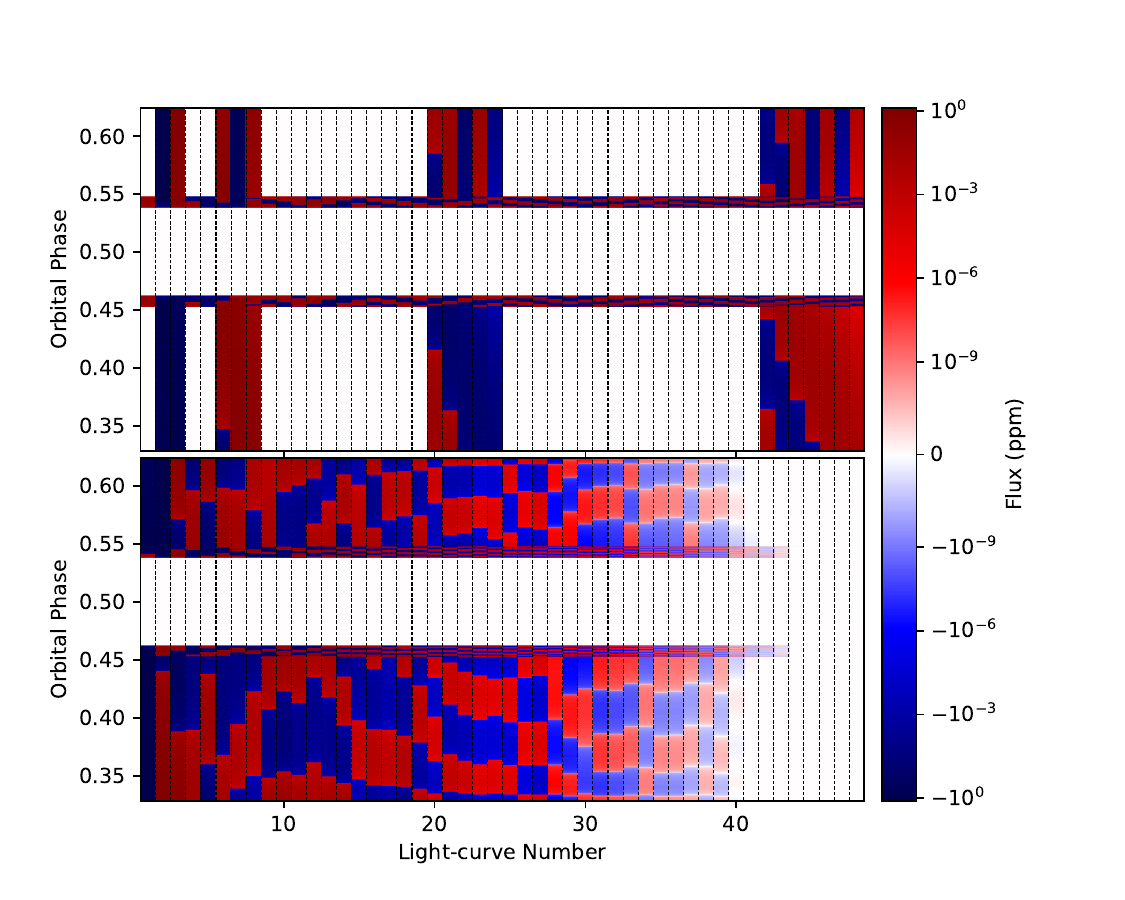}
    \caption{The spherical harmonic design matrix $\bm{\mathcal{A}}$ (top, same as the bottom panel of Figure \ref{fig:designmatrix}) and the eigencurve design matrix $\bm{\mathcal{A}_E}$ (bottom) for a JWST eclipse observation of WASP-18b \citep{CoulombeEtal2023arxivWASP18b}. $\bm{\mathcal{A}_E}$ is the result of orthogonalizing the light curves (columns) in $\bm{\mathcal{A}}$, and is the basis set of light-curve components used in eigenmapping fits. Both matrices have ranks of 43 (43 non-degenerate components), which is clear for $\bm{\mathcal{A}_E}$ where the orthogonalization has separated zero and nonzero components.}
    \label{fig:ecurves}
\end{figure*}

SVD sorts the eigencurves by their variance, naturally separating the observable components (nonzero variance) and null components (zero variance).
While the spherical harmonics design matrix for an eclipsed planet is non-zero for every harmonic mode (Figure \ref{fig:designmatrix}) but contains non-trivial degeneracies, the eigenmap design matrix has $R$ nonzero orthogonal components and $N - R$ null (degenerate) components. 
In other words, the ranks of the spherical harmonic design matrix and the eigenmap design matrix are equal.
This is evident in Figure \ref{fig:ecurves}, which shows $\bm{\mathcal{A}}$ with $0 < l \leq 6$ compared against $\bm{\mathcal{A}_E}$ for the WASP-18b eclipse observation in \cite{CoulombeEtal2023arxivWASP18b}.
Both matrices have $R = 43$, which is clear in $\bm{\mathcal{A}_E}$ where there are 43 nonzero, orthogonal columns.

The matrix $\bm{V}$ contains the eigenvectors which map from $\bm{\mathcal{A}}$ to $\bm{\mathcal{A}_E}$, as shown in Equation \ref{eqn:ae}.
As noted by \cite{RauscherEtal2018ajMap}, these eigenvectors, which are the weights to combine spherical harmonic light curves into eigencurves, are exactly the $\bm{y}$ values which produce the eigenmaps. 
Thus, we can calculate the physical map representations of the eigencurves following Equation \ref{eqn:map} and, because they are ranked by variance of the light curves they produce, inspect the last $N - R$ to further understand the null space.

\begin{figure*}
    \includegraphics[width=\textwidth]{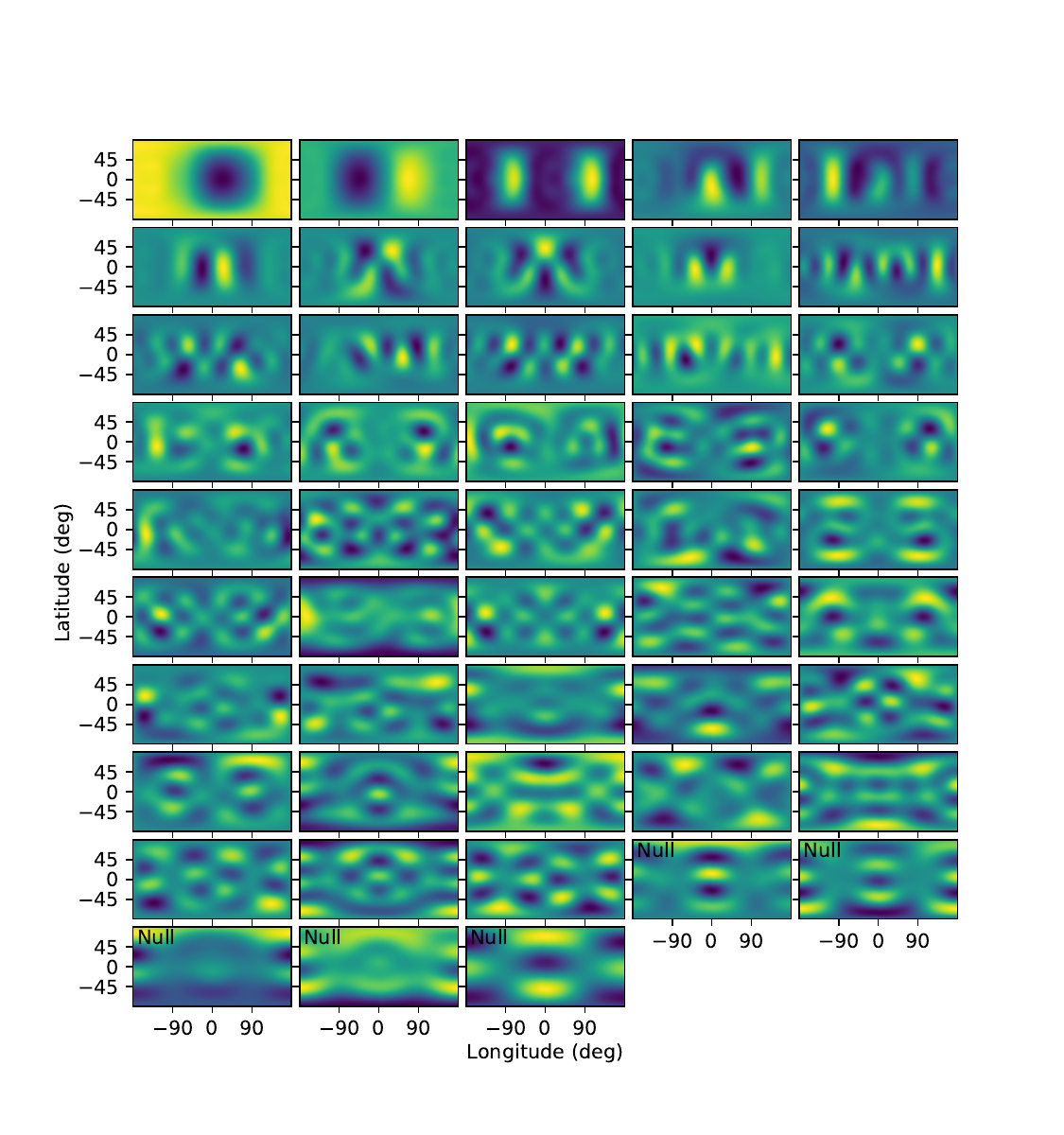}
    \caption{The flux (units arbitrary) eigenmaps corresponding to the eigencurves in a design matrix $\bm{\mathcal{A}_E}$ for $0 < l < 6$ for the eclipse observation of WASP-18b in \cite{CoulombeEtal2023arxivWASP18b}. These eigenmaps are ranked, from top left to bottom right, by the variance in their light curves. The final five maps generate light curves which are 0 at all times during the observation and, thus, are in the null space.}
    \label{fig:emaps}
\end{figure*}

Figure \ref{fig:emaps} shows all 48 eigenmaps corresponding to the eigencurves in $\bm{\mathcal{A}_E}$ for $0 < l < 6$. 
The final five eigenmaps are in the null space.
It is important to note that these null maps are model independent; they depend on the orbital parameters of the planet and observational settings (e.g., exposure time and orbital phase coverage) of the measurement, but their calculation requires no knowledge of the true planet map.
Thus, eigenmaps can be used to understand the null space of an observation before it is taken.

With eigenmapping, it is necessary to reduce the size of the model parameter space to prevent overfitting the data \citep{RauscherEtal2018ajMap, ChallenerRauscher2022ajThERESA, CoulombeEtal2023arxivWASP18b}. 
The fit is restricted to only $N_E$ of the highest-variance eigencurves for a given $l_{\rm max}$ (starting from the left in Figure \ref{fig:ecurves}), and fits are performed for many combinations of $N_E$ and $l_{\rm max}$, with the optimal choice selected as the combination which minimizes the Bayesian Information Criterion (BIC, \citealp{Raftery1995BIC}).
For reference, \cite{CoulombeEtal2023arxivWASP18b} found the optimal fit to a JWST white-light light curve is with $l_{\rm max} = 5$ and $N_E = 5$.
Thus, to avoid overfitting, many brightness patterns will not be found in the retrieved map, including the null space components.

Aside from the problem of overfitting, if the light-curve model is an unrestricted sum of eigencurves, it is not practical to include the null-space eigencurves in the fit. 
These eigencurves are exactly zero everywhere, so varying their weights does not affect the model light curve, so these weights will be completely unconstrained. 
However, the corresponding eigenmaps are not zero everywhere, so the resulting map will have large regions of infinite uncertainty. 
This is clearly not representative of reality, so the null eigenmaps (and other low-variance eigenmaps) have been ignored in eigenmap fitting, with the acknowledgment that retrieved maps and their uncertainties are limited by the components included in the fit \revision{\citep{RauscherEtal2018ajMap, ChallenerRauscher2022ajThERESA, CoulombeEtal2023arxivWASP18b}}.

However, the physically plausible parameter space is not a wholly unrestricted sum of eigencurves because the corresponding brightness map must be positive everywhere.
Therefore, recent eigenmapping applications \citep{ChallenerRauscher2022ajThERESA, CoulombeEtal2023arxivWASP18b} impose a positivity constraint, limiting the size of the model parameter space. 
Under this constraint, null-space eigencurves could be included in the fit without leading to unconstrained retrieved maps.

In the following sections we investigate several aspects of the null space:

\begin{enumerate}
    \item The impact of the null space on observations based on system parameters and observational settings, and strategies for mitigation.
    \item Using the null space to convert atmospheric forward models into null and observable components for better comparison with retrieved maps.
    \item The effects of including null-space components on the uncertainties of retrieved eclipse maps.
\end{enumerate}

\section{MITIGATING NULL-SPACE EFFECTS}
\label{sec:mit}

The null space is purely mathematical and, to a point, unavoidable.
However, through careful observational design and target selection, the impact of the null space can be minimized.
Note that here we only discuss considerations to minimize the nullity, and leave a discussion of optimizing eclipse mapping based on signal-to-noise considerations to future work.

\begin{figure}
    \centering
    \includegraphics{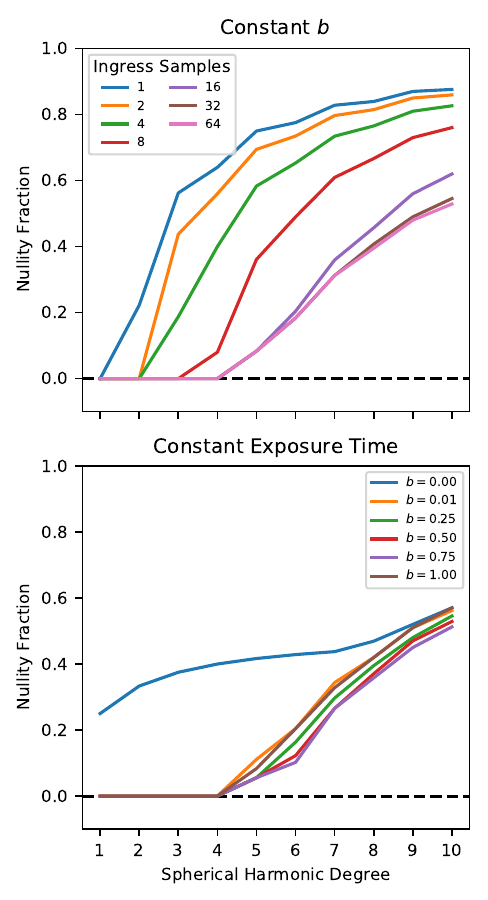}
    \caption{The size of the null space vs.\ exposure time (top) and impact parameter (bottom). The sampling rates range from 1 to 64 exposures over the eclipse ingress. \revision{The 64-sample case roughly matches the observation in \cite{CoulombeEtal2023arxivWASP18b}.} The impact parameters range from 0 to 1. In general, as long as there are $\geq 16$ exposures during eclipse ingress/egress (which will be standard for JWST), sampling rate does not affect the size of the null space. Similarly, the size of the null space does not change with impact parameter as long as the impact parameter is nonzero and the orbit is non-grazing.}
    \label{fig:obsdesign}
\end{figure}

The size of the null space is heavily influenced by the number of exposures occurring during eclipse ingress and egress. 
To demonstrate, we calculated the nullity for observations with varying numbers of exposures during ingress and egress, for a representative planet with $R_p/R_* = 0.1$, $M_p/M_* = 0.001$, an inclination of $89^\circ$, and a circular orbit with a period of 1 day.
Figure \ref{fig:obsdesign} shows the nullity as a fraction of the total number of harmonic modes as a function of ingress/egress samples for spherical harmonics of degree $\le10$.
For this representative planet, the duration of ingress/egress is 10.9 minutes, so our exposure times range from $\sim650$ - 10 seconds.
The nullity for this planet decreases dramatically, even for low-order harmonics, until reaching 16 exposures in both ingress and egress, corresponding to a sampling rate of $\sim40$ seconds\revision{; higher sampling rates do not further decrease the nullity for spherical harmonics of degree $\le5$}. 
This sampling rate \revision{limit} is much slower than the 8.86 seconds \revision{(approximately 64 ingress samples in Figure \ref{fig:obsdesign})} used in \cite{CoulombeEtal2023arxivWASP18b}, indicating that sampling rate will generally not affect the null space for JWST observations.
\revision{This also sets a limit for temporal binning in eclipse-mapping analyses, as binning decreases the spatial information content of the data and should be avoided when possible.}

As the durations of eclipse ingress and egress depend on impact parameter $b$, so too does the nullity, to a lesser degree than sampling rate.
Figure \ref{fig:obsdesign} shows the nullity for a range of impact parameters with a constant exposure time of 30 seconds. 
Note that increasing the impact parameter leads to a smaller nullity, until the impact parameter exceeds $1 - R_p/R_*$, when the eclipse becomes grazing, and the planet is only partially scanned by the stellar disk.
Similar effects (lower nullity, to a limit) are present for longer orbital periods and larger planetary radii, which both lead to longer ingress and egress durations.

\begin{figure}
    \centering
    \includegraphics{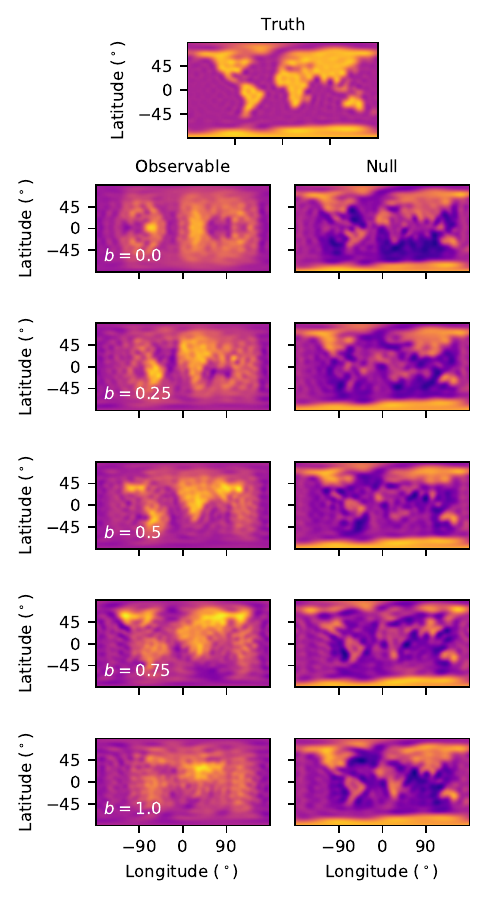}
    \caption{The observable (left) and null (right) modes for an eclipse of a land map of Earth (top) with a range of impact parameters. As expected, the edge-on ($b = 0$) and grazing ($b = 1$) have significant unobservable structures, and fine structures outside approximately -90 and 90$^\circ$ longitude are washed out. However, varying $b$ between these edge cases also changes the shape of the null space.}
    \label{fig:earth}
\end{figure}

Until now, we have discussed the size of the null space with respect to orbital parameters.
However, it is important to note that the shape of the null space can change dramatically for different orbital parameters, even if the overall nullity remains similar.
For example, varying the impact parameter within (0, $1 - R_p/R_*$] minimally changes the nullity (Figure \ref{fig:obsdesign}) but can significantly change which structures of the map are observable, particularly latitudinal variation.
To demonstrate, we computed observable and null modes, as described in Section \ref{sec:eigen} and further discussed below, for a well-sampled eclipse of a land map of Earth at several impact parameters, with other system parameters the same as the planet in Figure \ref{fig:obsdesign}; we chose this map because it has significant fine structure and latitudinal variation (Figure \ref{fig:earth}).
As discussed above, many structures are unobservable for the $b = 0$ and $b = 1$ cases, but there are also many differences in the null space among the other cases. 
Depending on the goals of an observation, it can be beneficial to consider the null space for the targets' orbital parameters.
We leave a full investigation into the interplay between orbital parameters, observational uncertainties, and the null space to future work.


\section{COMPARING THEORETICAL PLANET MODELS WITH RETRIEVED MAPS}
\label{sec:gcm}

A consequence of the null space is that an observed map may not match theoretical predictions from general circulation models (GCMs) of the planet.
GCMs are based on the physical processes of the atmosphere (e.g, radiation, winds, drag, etc.) and can generate brightness patterns which fall in the observational null space.
A retrieved map will not include these unmappable patterns.
Furthermore, as mentioned in Section \ref{sec:eigen}, the uncertainties on these retrieved maps are limited by the structures present in the components of the model fit, and the null space components are not included, so the map uncertainties will not encompass the effects of the null space.
Thus, there can be an apparent inconsistency between GCM forward models and retrieved maps from observations.



We propose that the correct approach is to transform a GCM flux map into its observable ($Z_\bullet$) and unobservable ($Z_\circ$) modes, through the $\bm{P}$ and $\bm{N}$ operators above. $Z_\bullet$ can then be directly compared with the retrieved planet map, and $Z_\circ$ shows the flux structures that are inaccessible.
First, one must represent the GCM flux map in spherical harmonics up to a high enough degree to capture spatial variations; in this work, we use 25th-degree harmonics.
Then, following L2021, compute $\bm{\mathcal{A}}$, decompose into $\bm{P}$ and $\bm{N}$, compute $\bm{y_\bullet}$ and $\bm{y_\circ}$, and use Equation \ref{eqn:map} to generate the maps.
In this work we perform these calculations using \texttt{starry} \citep{LugerEtal2019ajStarry}, which has built-in functionality for representing forward models in spherical harmonics and calculating spherical harmonic design matrices.
See Appendix \ref{app:code} for a  minimal code example of calculating observable and null maps using \texttt{starry}.

\begin{figure*}[t]
    \centering
    \includegraphics[width=7in]{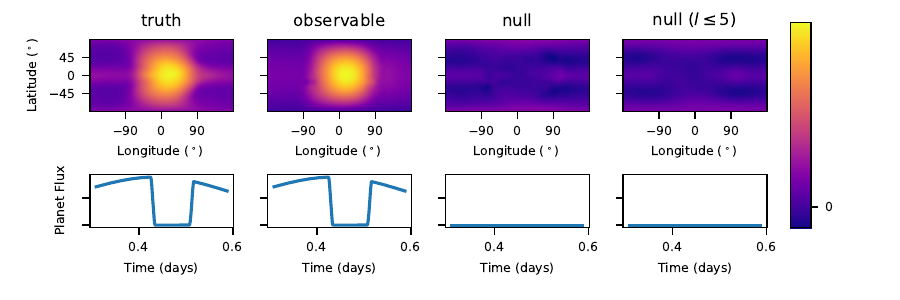}
    \caption{A decomposition of a GCM flux map into the observable modes and null space modes. The fourth column is a restriction of the full null space (third column) to only spherical harmonics of degree 5 or less, which shows the data-limited null space. Note that the light curve for the truth and the observable component are identical, and the light curves from the null space maps are flat, showing they are not observable.
    The GCM is the 0 G (no magnetic field) RM-GCM WASP-18b model from \cite{CoulombeEtal2023arxivWASP18b}.
    \label{fig:gcms}}
\end{figure*}

To demonstrate, we use a GCM of WASP-18b from \cite{CoulombeEtal2023arxivWASP18b}. 
We assume the same eclipse observation with the same observation times covering longitudes from -134.7$^\circ$ - 151.8$^\circ$.
Figure \ref{fig:gcms} shows a comparison between the true GCM map (represented with high-degree spherical harmonics), the observable modes, and the null space.
The light curves produced by the GCM map and its $Z_\bullet$ are identical, while the light curve produced by $Z_\circ$ is 0 flux at all times in the observation.

The most clear difference between $Z$ and $Z_\bullet$ is that outside the dayside of the planet, all temperature variations are washed out.
This is because, at longitudes that are not scanned by the stellar disk, latitudinal variation is unobservable and measurement of longitudinal variation is limited to large-scale structures. 
Thus, the relatively bright equator on the nightside due to a superrotating jet is undetectable.
Differences in $Z$ and $Z_\bullet$ on the dayside are more subtle.
Fine structure in the shape of the hotspot is lost, and the poles appear much dimmer than reality.


\section{EXAMPLE APPLICATIONS}
\label{sec:application}

To examine the consequences of the null space in some practical examples, we generated synthetic, noisy light curves based on GCMs.
We consider two hypothetical planets: an ultra-hot Jupiter similar to WASP-18b and a cooler planet similar to HD 209458b.
The GCM of WASP-18b is cloud-free, as the planet's atmosphere is too hot for condensation, and, thus, the temperature structure of the photosphere (i.e., what we are mapping) is primarily global-scale features with smooth horizontal gradients.
The HD 209458b GCM, however, has spatially-inhomogeneous cloud cover across the dayside, creating potentially-detectable (relatively) small-scale brightness structures.
These two planets represent two important temperature regimes which will be mappable with JWST.

For each planet, we consider two observational scenarios:

\begin{enumerate}
    \item \revision{Best-Case} Precision. This is an approximate best-case JWST observational scenario, assuming 10 stacked eclipses, each with 121 ppm precision per integration (matching \citealp{CoulombeEtal2023arxivWASP18b}) for a final precision of 38 ppm per integration\footnote{For eclipse-mapping purposes, reducing uncertainties by the square root of the number of eclipses is not identical to mapping with the additional eclipses, even for tidally-locked planets, because each \revision{exposure} will occur at slightly different orbital phases\revision{, changing (adding to) the information content of the light curve in a non-trivial way}. However, as shown in Section \ref{sec:mit}, additional sampling beyond a single JWST eclipse does not significantly affect the null space, so this noise approximation is reasonable for our purposes.}. This case shows how the null space impacts the highest-signal observations currently possible.

    \item Extreme Precision. These observations assume an extreme precision of 1 ppm to investigate the effects of the null space when mapping is less limited by data quality.
\end{enumerate}

\noindent
\revision{Both cases are relatively high-precision observations because single-eclipse observations are often satisfactorily fit by large-scale flux patterns (low-order harmonics), where the null space is empty.
That is, the information loss is more due to the strength of the eclipse-mapping signal rather than null-space limitation.
However, if a planet has a strong signal and flux patterns that cannot be fit by low-order harmonics, like the longitudinal temperature gradients in WASP-18b (see Section \ref{sec:w18}), then the null space can be important to consider for single-eclipse analyses.}

To generate the synthetic observations we take 3D temperature structures from the RM-GCM \citep{RauscherMenou2012apjGCM, RomanRauscher2017apjGCMupdate, BeltzEtal2022ajActiveMagDrag3D}. 
For WASP-18b we use the 0 G GCM from \cite{CoulombeEtal2023arxivWASP18b}.
For HD 209458b we use the 10\% condensation extended-clouds GCM with an irradiation temperature of 1500 K from \cite{RomanEtal2021apj3DClouds}.
System parameters are listed in Table \ref{tbl:syspar}.

\begin{table}[]
    \centering
    \begin{tabular}{l|r|r}
         Parameter & WASP-18 & HD 209458 \\
         \hline
         Stellar radius & 1.347 R$_\odot$ & 1.203 R$_\odot$\\
         Stellar mass & 1.5596 M$_\odot$ & 1.148 R$_\odot$\\
         Stellar temperature & 6400 K & 4371 K\\
         Planet radius & 0.132 $R_\odot$ & 0.139 $R_\odot$\\
         Planet mass & 0.00996 $M_\odot$ & 0.000702 $M_\odot$\\
         Eccentricity & 0.0 & 0.0\\
         Inclination & 84.38$^\circ$ & 86.59$^\circ$\\
         Impact parameter & 0.34 & 0.51\\
         Semimajor axis & 0.0218 AU & 0.04747 AU\\
         Orbital period & 0.94145 days & 3.52472 days\\
         \hline
    \end{tabular}
    \caption{WASP-18 \citep{CoulombeEtal2023arxivWASP18b} and HD 209458 \citep{RomanEtal2021apj3DClouds} System Parameters}
    \label{tbl:syspar}
\end{table}

We post-processed both GCMs following the methods of \cite{ChallenerRauscher2022ajThERESA} with minor changes. 
Briefly, we assume thermochemical equilibrium with GGchem \citep{WoitkeEtal2018aandaGGchem} and calculate spatially-dependent emission with Tau-REx \citep{Al-RefaieEtal2019arxivTauRExIII}.
For WASP-18b we calculated emission spectra over the NIRISS/SOSS first-order wavelength range (0.8 -- 2.8 \microns) and integrated to create a broadband measurement.
For the cloudy GCM, we assumed the 5 \microns\ cloud optical depths from the GCM and extended them to other wavelengths following the empirical relationship in \cite{LeeEtal2013apjHR8799b}, then integrated the resulting spectra over 5 -- 6 \microns.
Then, using \texttt{starry}, we transformed these flux grids to 25th degree spherical harmonics representations, which enables analytic computation of the eclipse light curve (i.e., we do not need to approximate the spatial integration as is done in \citealp{ChallenerRauscher2022ajThERESA}).

In this work we used ThERESA's \citep{ChallenerRauscher2022ajThERESA} 2D mapping mode to fit our data.
We test a large range of maximum spherical harmonic degree and number of eigencurves to use in the fit, selecting the optimal combination using the BIC.
\revision{Thus, we eliminate terms which, if they were fit, would be consistent with zero and would not significantly affect the best-fitting map.}
This is the same process used in \cite{CoulombeEtal2023arxivWASP18b} to produce the first JWST eclipse map.

\subsection{WASP-18b}

\begin{figure*}[t]
    \centering
    \includegraphics{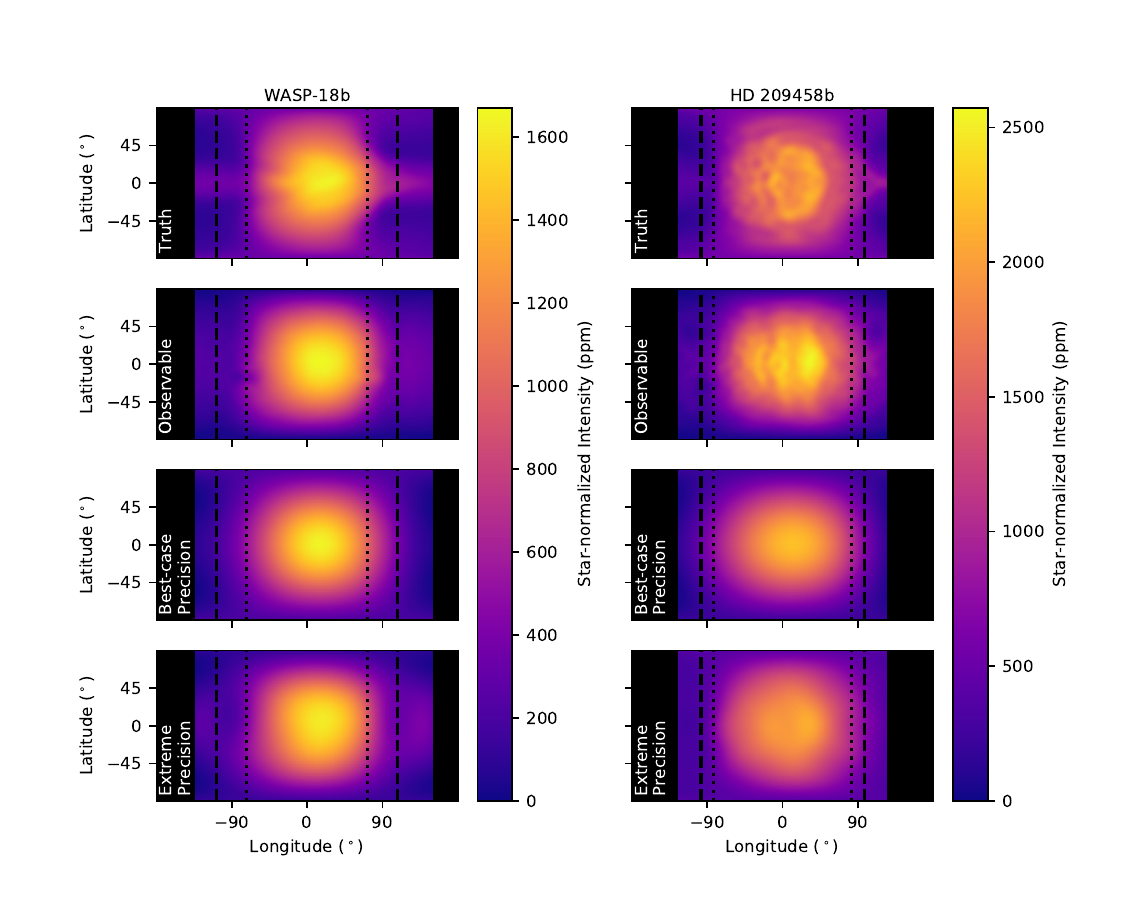}
    \caption{A comparison of the true planet brightness map, the measurable modes of the true map, the retrieved map from a fit to our \revision{best-case} precision synthetic data, and the retrieved map from a fit to our extreme precision synthetic data. The black bars cover regions of the planet that are not visible during the observation. The vertical dotted lines bound the longitudes of the planet which are scanned during both ingress and egress. Within the dashed lines are longitudes scanned by at least one of ingress or egress. Outside the dashed lines, all information comes from only the phase-curve variation.}
    \label{fig:fit}
\end{figure*}

The fitted maps of WASP-18b are shown in Figure \ref{fig:fit}, compared against the truth and the observable modes of the truth.
\revision{The best-case precision data require $l_{\rm max} = 3$ and $N_E = 2$, while the extreme-precision case requires $\l_{\rm max} = 5$ and $N_E = 13$}.
For large-scale features, like the location and extent of the planet's hotspot, we achieve excellent agreement between both fitted maps and the true map. 
As expected, the agreement is particularly good at longitudes scanned by the stellar disk during both ingress and egress.

Some smaller-scale features of the true map are not recovered.
The finer structure of the western edge of the hotspot, despite being well within the longitudes probed by the eclipse, are not found in our retrieved maps. 
At the terminators, the bright equator and dimmer mid-latitudes are not present in our fit.
For the \revision{best-case} precision case, we recover a much dimmer nightside than the truth, but a mismatch here is to be expected as the nightside is very poorly constrained by an eclipse observation, and the retrieved map only needs to be consistent with the data on a hemispheric scale outside the dashed lines.
As expected, the fit to the extreme precision case achieves a more accurate average nightside temperature, although limitations from the null space still prevent retrieval of spatial structures in that region.

However, the retrieved maps match the observable modes of the true map remarkably well, particularly at the longitudes of the planet which are probed by the eclipse. 
Despite differences between the fitted maps and the truth, even in the extreme precision case, we should consider these maps to be consistent with this GCM prediction.
That is, in a real scenario where we retrieved this map, not knowing the truth, and compared the fitted map with this GCM forward model, one might be concerned with the differences between them.
If the GCM is transformed to its observable component, however, we see that the GCM actually matches the observation well.\footnote{Here, the observable component of the GCM and the fitted map are guaranteed to match because the GCM was used to generate the data. Regardless, this exercise shows how to make appropriate comparisons between forward models and fitted maps even when the truth is unknown.}

For this planet, the \revision{best-case} precision map and the extreme precision map are very similar.
The minimal differences are primarily in the nightside brightness and the latitudinal extent of the hotspot.
The nightside brightness is highly uncertain due to limited observing time during typical eclipse observations, so the much lower light-curve uncertainties of the extreme precision case lead to a more accurate nightside measurement.
The signature of latitudinal variation in eclipse ingress and egress is very small due to WASP-18b's low impact parameter ($\approx 0.34$) so constraints on latitudinal information are minimal without very high observational precision.
This can be seen in that the features in the null space are primarily latitudinal (Figures \ref{fig:emaps} and \ref{fig:gcms}).

We note that, although these differences in the best fits between these two cases are minor, achieving higher-signal observations is still beneficial to mapping.
The difference in precision between the \revision{best-case} precision and extreme precision maps is significant, which has implications for the measurement of properties like longitudinal hotspot offsets, latitudinal hotspot offsets, and spatial temperature gradients.
We discuss this in further detail in Section \ref{sec:hotspots}.

\subsection{HD 209458b}

Figure \ref{fig:fit} shows the true map, observable modes, and fits to the synthetic data of HD 209458b.
\revision{The best-case precision data require $l_{\rm max} = 3$ and $N_E = 3$, while the extreme-precision case requires $\l_{\rm max} = 8$ and $N_E = 18$}.
In contrast to WASP-18b, this planet has significant fine spatial brightness variation due to patchy clouds causing the photospheric pressure (and, thus, temperature) to change with latitude and longitude.
That is, cloudy regions are more optically thick so an observation probes lower, colder pressures in those regions.

There is still evidence of cloud patchiness on the dayside in the observable modes of the GCM but much of the precise structure is unobservable, even in eclipse.
There are larger differences in brightness across the dayside of the observable map than the truth, with brighter regions near the substellar point and dimmer poles.
The observable hotspot is confined to a smaller range of latitudes and, as with WASP-18b, most spatial information on the nightside is lost.

For the \revision{best-case} precision case, the retrieved map shows none of the substructure associated with patchy clouds on the dayside.
The data are well fit by a model with a smooth brightness distribution decreasing from a hotspot east of the substellar point toward the terminators and poles. 
Broadly, the shape of the hotspot matches the shape of the observable map hotspot.
With the extreme precision case, we see some substructure from patchy clouds, with two brighter regions east and west of the substellar point.

\subsection{Measuring Hotspot Offsets with Eclipse Mapping}
\label{sec:hotspots}

Hotspot offsets are often measured to give insight into atmospheric circulation patterns \citep[e.g.,][]{StevensonEtal2017ajWASP43bPhaseCurve, KreidbergEtal2019natLHS3844b, MorelloEtal2019ajW43bPhaseCurves, CrossfieldEtal2020apjlLTT9779bPhaseCurve, MayEtal2021ajWASP76bPhaseCurves, DelineEtal2022aapW189CHEOPSphasecurve, MayEtal2022ajSpitzerPhaseCurves, MercierEtal2022aj55CncePhaseCurve}.
With phase-curve observations, the hotspot offset is measured as the orbital phase at which the planet reaches maximum emission; that is, the hotspot offset is really the planetary rotation angle which creates the brightest possible visible hemisphere. 
Eclipse mapping provides maps at much finer resolution than hemisphere-scale features such that a derivation of the hotspot location is the actual latitude and longitude of the hottest point on the planet.
These measurements can differ at a statistically-significant level, so authors must be careful to mention which measurement they are making \citep{ParmentierEtal2021mnrasCloudyPhaseCurves}.
Here, we refer to ``phase-curve offsets'' (planet rotation angle of maximum hemispheric brightness) and ``hotspot longitudes'' (longitude of maximum brightness).

\begin{figure}
    \centering
    \includegraphics{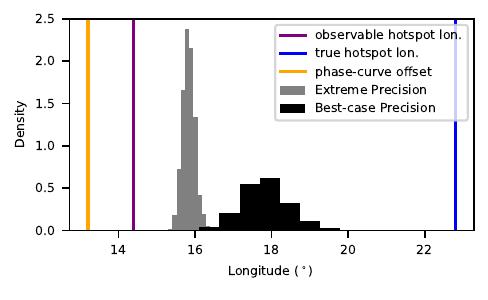}
    \caption{Hotspot longitudes of the true map and observable map of WASP-18b compared against the retrieved hotspot longitudes and the phase-curve offset. Vertical lines denote the hotspot longitudes and the phase-curve offset. Note the difference in hotspot longitudes between the true map and the observable map. In black are the histograms of retrieved hotspot longitudes for the \revision{Best-Case} Precision and Extreme Precision cases showing evidence of a bias in the measured hotspot location compared to both the true location and the location we expect to observe.}
    \label{fig:hotspot}
\end{figure}

To demonstrate, we examined the hotspot longitude of the WASP-18b true map, observable map, retrieved maps, and phase-curve offset, shown in Figure \ref{fig:hotspot}.
The true and observable hotspot longitudes are computed by generating high-resolution maps and finding the location of maximum brightness.
The phase-curve offset is calculated by finding the orbital phase that achieves maximum planet brightness, assuming the planet is not occulted by the star.
The retrieved hotspot longitude histograms come from the longitude of maximum brightness for a subsample of the ThERESA MCMC results.

We note that the phase-curve offset is nearly 10$^\circ$ west of the hotspot longitude of the true map.
This is because the dayside flux distribution is longitudinally asymmetric (Figure \ref{fig:hotspot}, top panel), with a steeper gradient at the evening terminator, so the hemisphere of maximum brightness is shifted west relative to the true hotspot location.
These are different measurements, as described above, but we mention the difference to highlight that eclipse-map hotspot longitudes and phase-curve offsets will not necessarily be consistent with each other.

Furthermore, the observable map hotspot longitude and the retrieved map hotspot longitudes are also west of the true hotspot longitude. 
While with increasing observational precision the measured hotspot longitudes approach the observable map hotspot longitude (what we expect to measure with infinite S/N), even at an extreme precision the retrieved hotspot longitude is biased west of the true hotspot longitude and east of the observable hotspot longitude.
This is likely because the exact location of the hottest point on the planet is relatively unimportant to the eclipse-mapping fits compared to the overall dayside flux morphology.
With upcoming eclipse-mapping analyses we must be aware that measured hotspot longitudes may not match both GCM predictions and the observable transformations of those predictions. 
Using eclipse-mapping hotspot longitudes as a single metric to understand atmospheric dynamics will be challenging, and methods which utilize the full shape of the planet map may be necessary \citep{HammondLewis2021pnasRotDivMap}.
\cite{CoulombeEtal2023arxivWASP18b} showed that the traditional hotspot longitude becomes less meaningful with eclipse maps, and the problem only gets worse for planets with complex dayside structures like the map of HD 209458b presented here.




\section{INCORPORATING NULL-SPACE UNCERTAINTY IN ECLIPSE MAPPING}
\label{sec:unc}

As described above, if null-space light curves are included in an unrestricted light-curve fit (i.e., the model is simply a sum of light curves) then the null-space components are unconstrained and the resulting map will have infinite uncertainty.
However, we can place physically-motivated penalties on the model which restrict parameter space, potentially allowing the inclusion of null-space modes without dramatically inflating map uncertainties.
For example, we can impose a positivity constraint which significantly penalizes goodness-of-fit for models that result in negative fluxes on regions of the planet that are visible during the observation.
This constraint is already available in ThERESA.

Here, we investigate whether the uncertainties introduced by the null space are manageable under this positivity constraint.
For simplicity, we utilize the same BIC-based model optimization procedure described in \cite{ChallenerRauscher2022ajThERESA} to choose $l_{\rm max}$ and $N_E$.

Given this best-ftting model, one could then analytically calculate the largest magnitude positive and negative weights on the null components that keep the best-fitting brightness map positive in the visible regions of the planet.
If the null components are uncorrelated with other model parameters then this would give the range of plausible models.
However, while the eigencurves and null-space components are, by design, orthogonal in the light curves they produce, the positivity constraint removes regions of parameter space, introducing correlations.
Additionally, the best-fitting model is only one of a range of possible solutions, all of which may interact differently with the null-space components.

Therefore, in order to fully explore the effects of the null-space components on the fit, we rerun the MCMC using the same $l_{\rm max}$ and $N_E$ but including the null-space components present at this $l_{\rm max}$.
This does not change the best-fitting model (the null-space components provide no change to the light-curve model), but does allow the model the explore a larger range of brightness maps.

We calculate brightness map uncertainties by evaluating maps from the MCMC posterior distribution and calculating 68.27, 95.45, and 99.73\% confidence regions at each latitude/longitude in this map distribution.
Thus, these uncertainties are the range of possible values the map can take at a specific location, but not every possible map within a confidence region is a potential match to the data.
While the eigencurves are, by definition, orthogonal in light-curve space, their corresponding eigenmaps are not orthogonal so there can be correlations in the posterior distribution of maps.
For example, if a map has very bright regions it likely also has very dim regions to maintain a total emission that is consistent with the data.

\subsection{Application to Synthetic Data Sets}

We applied this methodology to the extreme precision datasets described above, as the \revision{best-case} precision data are explained well by eigencurves with $l_{\rm max} \leq 3$ where the null space is empty.
\revision{This means that, for these specific GCMs, the null-space uncertainty method outlined here, when used in tandem with the BIC-based model-selection approach, would not alter the mapping uncertainties for the best-case (or lower) precision synthetic data.
However, this is not a guarantee that real data at lower precision will not benefit from this null-space uncertainty analysis; real planets will likely have brightness components not precisely the same as those present in these GCMs.
For example, the WASP-18b data in \cite{CoulombeEtal2023arxivWASP18b} consist of only one eclipse but require $l_{\rm max} \leq 5$ to be fit, such that the null space is not empty (see the following section).
}

\begin{figure*}
    \includegraphics[width=\textwidth]{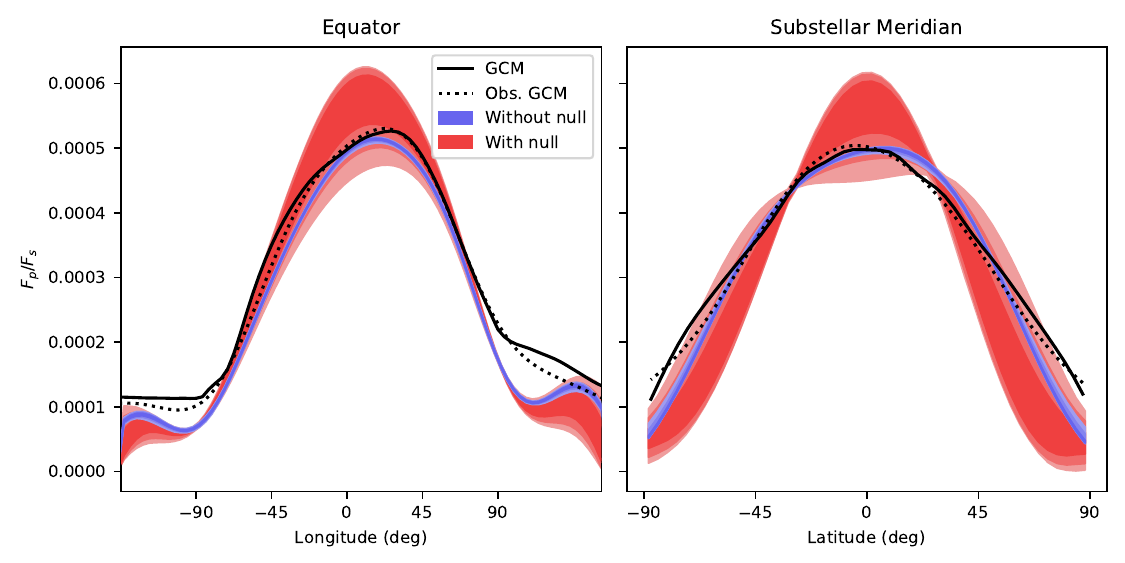}
    \caption{Uncertainties (68.27, 95.45, and 99.73\% quantiles) of fits to the synthetic WASP-18b extreme precision eclipse observation with (red) and without (blue) null-space components in the model. The GCM ground truth and observable component are overplotted. The model without the null space captures the large-scale brightness variations but the uncertainties are greatly underestimated and do not encompass the ground truth. The model with the null space has significantly increased uncertainties which encompass the truth across much of the dayside, where the observation is sensitive to two-dimensional spatial information.}
    \label{fig:w18-synth-null}
\end{figure*}

The WASP-18b \revision{extreme precision} synthetic light curve is best fit with $l_{\rm max}$ = 5, where there is one null space component.
Figure \ref{fig:w18-synth-null} compares the WASP-18b brightness-map uncertainties between fits with and without the null-space components along the equator and the substellar meridian, with the true brightness map from the GCM and its observable component overplotted.
Here, due to the extreme precision of the light curve, the model without null components has extremely low uncertainties.
Once we include the null-space component, which is unconstrained by the observational uncertainties, the range of plausible models increases significantly.

The GCM, which is the ground truth, does not fall within the uncertainties of the model fit without the null-space components, which is expected (see Section \ref{sec:gcm}).
In fact, the uncertainties on the fit without the null component do not encompass the observable component of the GCM, although the fit is a closer match to the observable component of the GCM than the unprocessed GCM.
This is because the observable component of the GCM is calculated using spherical harmonics up to 25th degree while the fit uses $l_{\rm max} = 5$, so the fit is incapable of capturing some of the smaller scale features and sharper gradients in the observable GCM.
Given infinite S/N, in which case we would be justified in fitting the observation with a large number of eigencurves, the fit without the null components would precisely match the observable component of the GCM.
If we include the null space component in the fit, the increased uncertainties encompass the observable GCM \revision{and the ground truth} over most of the dayside of the planet, particularly near the substellar point and at low latitudes, where the visibility function is highest (lowest angle between the observer and location on the planet) and the planet is observed throughout the eclipse observation.
That is, the uncertainties are most accurate at locations to which the observation is most sensitive.

\begin{figure*}
    \includegraphics[width=\textwidth]{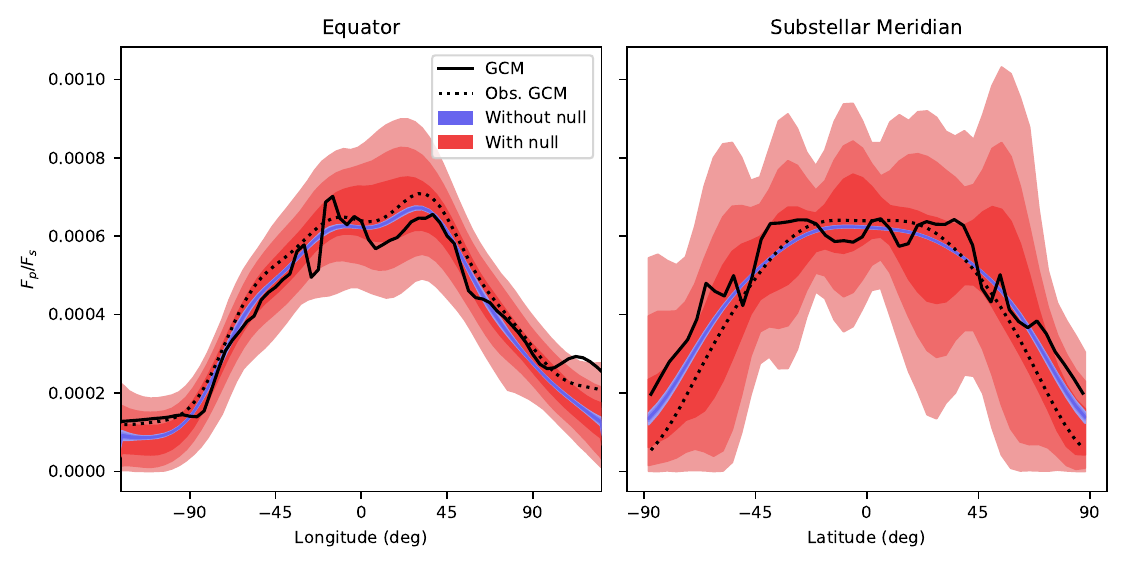}
    \caption{Same as Figure \ref{fig:w18-synth-null} but for the synthetic extreme-precision eclipse observation of HD 209458 b. When the null components are included in the model exploration, the model uncertainties encompass the observable modes of the GCM and the unprocessed GCM.}
    \label{fig:hd209-synth-null}
\end{figure*}

Because the HD 209458 b GCM has significant fine spatial brightness variations, fitting the extreme precision synthetic light curve requires $l_{\rm max} = 8$ and $N_E = 18$.
This has two noticeable effects compared to WASP-18b: 1) the larger $N_E$ gives the fit additional flexibility which allows the fit without null components to closely match the observable component of the GCM, and 2) at this higher spatial precision the null space contains 27 components, and including all these components causes enormous uncertainty inflation on the retrieved map (Figure \ref{fig:hd209-synth-null}).

These large uncertainties could be the reality of fitting fine spatial structures with eclipse mapping in the face of unobservable brightness patterns.
However, the match between the GCM ground truth is well within the 95.45\% confidence region of the fit, and at most locations within the 68.27\% confidence region, possibly indicating that we are overestimating our uncertainties with this methodology.
There are potentially other physically-motivated constraints we could place on the brightness map to further restrict parameter space.
For example, a lower limit of zero on the flux is quite conservative; these planets are very hot, so we could be justified in using a higher limit based on expected atmospheric temperatures. 
Such considerations may be important as eclipse-mapping data quality improves in the future.

\subsection{Application to WASP-18b JWST Eclipse}
\label{sec:w18}

For an example application to real data, we examined how including null space components in a fit to a WASP-18b JWST NIRISS/SOSS eclipse observation \citep{CoulombeEtal2023arxivWASP18b} affected uncertainties on the map.
For this observation, the BIC-optimized fit uses $l_{\rm max} = 5$.\footnote{\cite{CoulombeEtal2023arxivWASP18b} note that they achieve a similarly good fit with $l_{\rm max} = 2$ and $N_E = 5$. At this low harmonic degree there are no null-space components, so following the procedure we outline here will not change the uncertainties on that fit.}
At this harmonic degree there is one eigencurve/eigenmap in the null space.

\begin{figure}
    \centering
    \includegraphics[width=0.5\textwidth]{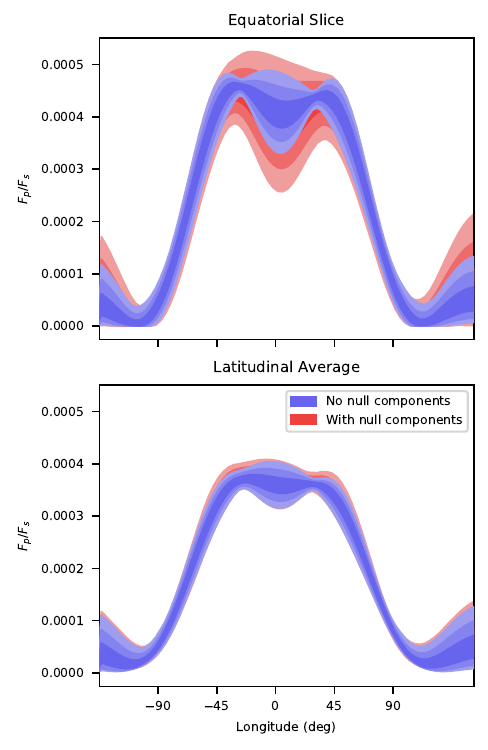}
    \caption{A comparison between the uncertainties on the WASP-18b eclipse map from \cite{CoulombeEtal2023arxivWASP18b} (blue) and a fit that is allowed explore the null space (red). Regions denote the 68.27, 95.45, and 99.73\% confidence regions. As expected, adding the null space component to the fit increases the uncertainty but, due to the positivity constraint, the uncertainties are kept finite. \textbf{Top:} An equatorial slice of the maps. \textbf{Bottom:} A $\cos{\rm (latitude)}$-weighted brightness map. The null space primarily introduces latitudinal variations that inflate uncertainties at individual locations on the map but the effect on the average longitudinal structure is minimal.}
    \label{fig:w18-null}
\end{figure}

At individual locations on the map, allowing the fit to explore the null space increases uncertainties by up to a factor of $\approx2$ (Figure \ref{fig:w18-null}, top panel). 
At the terminators, the uncertainties are nearly identical between the two fits, but the fit with the null component has larger uncertainties across the dayside and beyond the terminators.
These regions correspond with the brightness patterns in the null component, which has bright poles and a dim equator on the dayside, and vice versa on the nightside.
If we average over the latitudinal variations in the map, using a $\cos{\rm (latitude)}$ weighting as is done in \cite{CoulombeEtal2023arxivWASP18b}, the uncertainty introduced by the null component is significantly reduced (\ref{fig:w18-null}, bottom panel).
Thus, longitudinal brightness patterns can be measured robustly from a single JWST eclipse, even when incorporating the uncertainty from unobservable patterns.

This increase in map uncertainties does not change the scientific conclusions of \cite{CoulombeEtal2023arxivWASP18b}.
The evidences for atmospheric drag -- sharp temperature gradients moving from the substellar point to the terminators and a lack of a hotspot offset -- remain.
Likewise, there is still a preference for a dayside ``plateau'' over a traditional hotspot.
However, these uncertainties are more realistic as they incorporate brightness patterns that are unobservable but potentially present in the planet's atmosphere.
These new uncertainties are much more intuitive, as they do not require caveats about the null space (for harmonics less than or equal to the harmonic degree used in the fit).

\revision{We recommend that future eigenmapping analyses include the effects of the null space on map uncertainties in this manner. To summarize:}

\begin{enumerate}
    \item \revision{Use the BIC to optimize for $l_{\rm max}$ and $N_E$.}
    \item \revision{Determine which eigenmaps at the optimal $l_{\rm max}$ are in the null space following Section \ref{sec:eigen}.}
    \item \revision{Re-explore parameter space for the optimized model while including the null-space eigenmaps at the optimized $l_{\rm max}$ alongside the $N_E$ eigenmaps used in step 1.}
\end{enumerate}

\noindent
\revision{Thus, if the data warrant spatial variations at high spatial resolution (high $l_{\rm max}$), the mapping uncertainties will include the effects of the null space at that spatial resolution.}

\section{CONCLUSIONS}
\label{sec:conclusions}

With JWST we can use eclipse observations of exoplanets to retrieve brightness and temperature maps of their atmospheres.
These maps will provide crucial insight into the physical processes at work and will be a critical test of theoretical predictions of exoplanet atmospheres.
However, some brightness patterns are inaccessible to eclipse mapping which can complicate interpretations.

We described the null space and showed how the number of null components (the nullity, or size of the null space) depends on both orbital parameters and observational settings.
Typical JWST exoplanet eclipse observation exposure times are short enough to minimize the size of the null space. 
We noted that the shape of the null space -- which structures of a given map are inaccessible to eclipse mapping -- changes with orbital parameters, so depending on the science goals of an observation, careful target selection can limit the impact of the null space.

The null space can create differences between retrieved maps and GCMs that, at first, appear to raise questions about the GCMs' or eclipse maps' accuracy.
We have shown that if the GCMs are first transformed into the brightness patterns which are accessible to eclipse mapping, they provide a much closer match to retrieved maps.
To demonstrate this, we generated synthetic eclipse light curves of WASP-18b and HD 209458b from clear and cloudy GCMs, respectively, under two observational scenarios: \revision{best-case} precision mapping and extreme precision mapping.
We then fit eclipse maps to these light curves, comparing them against the truth and the observable modes of the GCMs. 

For the cloudless ultra-hot Jupiter WASP-18b, the GCM ground truth shows an eastward-shifted hotspot shaped by a superrotating equatorial jet.
Maps fit to both the \revision{best-case} precision and extreme precision cases recover the eastward-shifted hotspot but are unable to recover the smaller-scale features.
This matches the observable modes of the GCM quite well, indicating that while there are differences between the true map and the retrieved maps, in fact the retrieved maps are consistent with the GCM.

These differences become more apparent in planets with smaller-scale spatial inhomogeneities, as shown by our application to a cloudy model of HD 209458b.
The observable modes of the GCM lack much of the fine spatial structure present in the true map, although there is still evidence of the brightness variations introduced by the clouds.
The \revision{best-case} precision map does not retrieve these small-scale variations, but does match the large-scale features of the observable modes.
The extreme precision map shows some cloud-like flux variations, but the variations are weak due to a combination of the null space reducing visible inhomogeneities and the weakness of the signal generated by those inhomogeneitites (in principle, with infinite S/N, we would recover exactly the observable map, which clearly shows the small-scale effects of the clouds).

Finally, we presented a method to incorporate the uncertainty introduced by the null space.
By definition, null map components are unobservable and, thus, not constrained by observations.
However, if we place a positive-flux constraint on map models, we can limit model parameter space such that the range of plausible map models is finite even when including null components.
We demonstrated this approach with synthetic data, showing that the updated uncertainties on the fitted maps encompass the ground truth maps used to generate the data. 
When applied to a real JWST eclipse, this method primarily increased latitudinal uncertainty, but conclusions based on inferred longitudinal structure remained unchanged.

JWST and future telescopes offer an exciting new opportunity to understand the multidimensional properties of exoplanet atmospheres through eclipse mapping.
Our interpretations of exoplanet maps will be driven by comparison with theoretical predictions from GCMs, and advancements in GCMs will be driven by the maps we observe.
It is critical that we understand how retrieved maps are affected by the null space, and that we convert theoretical predictions to their observable modes before comparing them to observations.

\begin{acknowledgments}

  \revision{The authors thank the anonymous referee for their constructive comments, which improved the quality of this manuscript.}
  We thank Rodrigo Luger and Fran Bartolic for discussions on
  the null space.
  We thank contributors to SciPy, Matplotlib, Numpy, and the Python
  Programming Language. 
  This work was supported by a grant from the Research Corporation for Science Advancement, through their Cottrell Scholar Award.
\end{acknowledgments}

\software{NumPy \citep{HarrisEtal2020natNumPy}, Matplotlib
  \citep{Hunter2007cseMatplotlib}, SciPy
  \citep{VirtanenEtal2020natmSciPy}, Scikit-learn,
  \citep{PedregosaEtal2011jmlrScikitLearn}, starry
  \citep{LugerEtal2019ajStarry}, ThERESA \citep{ChallenerRauscher2022ajThERESA}}

\appendix

\section{Calculating Observable and Null Maps from Forward Models}
\label{app:code}

Here we show a minimal example of how to calculate the observable and null maps of a forward model using \texttt{starry}.
First, define the \texttt{starry} objects based on the system parameters and load the forward model (\texttt{gcm}, here) to calculate a spherical-harmonic representation of the forward model:

\begin{verbatim}
    import starry
    import numpy as np

    # Turn off starry lazy calculation (see documentation)
    starry.config.lazy = False

    # Set system paramters
    t0 = 0.0 # transit time (days)
    porb = 1.0 # orbital period (days)
    rs = 1.0 # stellar radius (solar radii)
    rp = 0.1 # planet radius (solar radii)
    ms = 1.0 # stellar mass (solar masses)
    mp = 0.001 # planet mass (solar masses)

    # Define starry objects and load an input flux map
    pmap = starry.Map(ydeg=25)
    smap = starry.Map(ydeg=1)

    pmap.load(gcm)
    yval = np.copy(pmap.y)

    planet = starry.Secondary(pmap, porb=porb, m=mp, r=rp)
    star   = starry.Primary(smap, m=ms)

    system = starry.System(star, planet)  
\end{verbatim}

Next, calculate the position and size of the occulter (star) relative to the planet, in units of the planet's radius, at the times in the observation.
This is straightforward with \texttt{starry}:

\begin{verbatim}
    # Define exposures and planet phase
    time = np.linspace(0.4, 0.6, 1000)
    theta = 360. * time / porb

    # Calculate planet-star relative positions and size
    x, y, z = system.position(time)

    xo = (x[0] - x[1]) / rp
    yo = (y[0] - y[1]) / rp
    zo = (z[0] - z[1]) / rp   
    ro = rs / rp
\end{verbatim}

Finally, calculate the design matrix for this observational setup, removing the $Y^0_0$ term as discussed in the main text, and calculate the $\bm{\mathcal{P}}$ and $\bm{\mathcal{N}}$ operators:

\begin{verbatim}
    # Calculate the design matrix
    A = pmap.design_matrix(theta=theta, xo=xo, yo=yo, zo=zo, ro=ro)[:,1:]

    # Calculate the operators (Equations 8 and 9)
    rank = np.linalg.matrix_rank(A)
    U, S, VT = np.linalg.svd(A)
    N = VT[rank:].T @ VT[rank:]
    P = VT[:rank].T @ VT[:rank]    
\end{verbatim}

One can apply these operators to the spherical-harmonic representation of the forward model and calculate the resulting maps using \texttt{starry}:

\begin{verbatim}
    # Set the spherical harmonic weights for the null map and calculate
    pmap[1:,:] = N @ yval[1:]
    nullmap = pmap.render()
    nullmap -= pmap.amp / np.pi # Subtract the uniform component

    # Set the spherical harmonic weights for the observable map and calculate
    pmap[1:,:] = P @ yval[1:]
    obsmap = pmap.render()
\end{verbatim}

\revision{Note that for visualization purposes we have subtracted the uniform component from the null map, such that the sum of the null and observable maps will equal the GCM. 
In reality, we have computed the observable and null maps while ignoring the uniform component, as we are interested in deviations from a uniform map, so one could also subtract the same uniform component from the GCM and the observable map.}






\bibliography{nullspace.bib}

\end{document}